\documentclass{optica-article}

\journal{opticajournal} 

\articletype{Research Article}
\usepackage{bm}
\usepackage{xspace}
\usepackage{amsmath}
\usepackage{graphicx}
\usepackage{xcolor}
\usepackage{comment}
\usepackage{multirow}
\usepackage{rotating}
\usepackage{diagbox}
\usepackage{subcaption}
\usepackage{placeins}
\usepackage{float}
\usepackage{booktabs}

\graphicspath{{./Figures}{.}}
\newcommand{\etal}{\textit{et al.\@}\xspace}
\newcommand{\exvivo}{\textit{ex vivo}\xspace}

\newcommand{\invivo}{\textit{in vivo}\xspace}

\newcommand{\invitro}{\textit{in vitro}\xspace}

\newcommand{\Enface}{\textit{En face}\xspace}
\newcommand{\enface}{\textit{en face}\xspace}

\newcommand{\enfaceh}{\textit{en-face}\xspace}
\newcommand{\um}{\(\muup\)m\xspace}
\newcommand{\uM}{\(\muup\)M\xspace}
\newcommand{\bmx}{\bm{x}}     \newcommand{\bmxp}{\bm{x}'}   \newcommand{\bmxpp}{\bm{x}_p} \newcommand{\bmxqq}{\bm{x}_q} \newcommand{\deltax}{\Delta\bm{x}} \newcommand{\dd}{\bm{\mathrm{d}}}
\newcommand{\rbra}[1]{{\left({#1}\right)}}
\newcommand{\cbra}[1]{{\left\{{#1}\right\}}}
\newcommand{\abra}[1]{{\left[{#1}\right]}}

\newcommand{\RN}[1]{\uppercase\expandafter{\romannumeral #1}}

\newcommand{\memo}[1]{\xspace}

\begin{document}
\title{Numerical speckle reduction for optical coherence tomography based on an analytical image formation model
}

\author{Xibo Wang,\authormark{1}
	Shuichi Makita,\authormark{1}
	Nobuhisa Tateno,\authormark{1}
	Suzuyo Komeda,\authormark{1}
	Cunyou Bao,\authormark{1}
	Atsuko Furukawa,\authormark{2}
	Satoshi Matsusaka,\authormark{2}
	Makoto Kobayashi,\authormark{3}
	and Yoshiaki Yasuno\authormark{1,*}}
\address{\authormark{1}Computational Optics Group, University of Tsukuba, Tennodai 1-1-1, Tsukuba, Ibaraki, Japan.\\
	\authormark{2}Clinical Research and Regional Innovation, Institute of Medicine, University of Tsukuba, Tennodai 1-1-1, Tsukuba, Ibaraki, Japan.\\
	\authormark{3}Department of Molecular and Developmental Biology, Institute of Medicine, University of Tsukuba, Tennodai 1-1-1, Tsukuba, Ibaraki, Japan.}

\email{\authormark{*}yoshiaki.yasuno@cog-labs.org}

\begin{abstract*}
	Speckle is an intrinsic granular pattern that limits the effective resolution of optical coherence tomography (OCT).
	We propose a numerical speckle reduction method using the ``real part of a shifted complex conjugate product'' (real-SCCP), which is a new image representation designed based on an analytical imaging model of OCT.
	In the real-SCCP image representation, the speckle can be fully numerically modulated after OCT signal acquisition.
	By averaging multiple SCCP images with different speckle realizations, the speckle is effectively reduced.
	Furthermore, this method enables clear visualization of structures in tumor spheroids and zebrafish eyes, and significantly outperforms conventional frame-averaging-based speckle reduction techniques.
\end{abstract*}

\section{Introduction}
\label{sec:Introduction}
Recent improvements in cell culture techniques have enabled three-dimensional (3D) \invitro models such as organoids and spheroids \cite{gunti2021organoid,el2023cancer}, which have been widely adopted in biomedical and life science research.
Optical coherence tomography (OCT) \cite{huang1991optical} is expected to be suitable for non-invasively investigating such thick 3D cell cultures.
However, the lateral resolution of OCT is conventionally configured to be relatively low, such as around 5 to 10 micrometers, due to the trade-off between lateral resolution and depth of focus (DOF).

Full-field swept-source OCT (FF-SSOCT) \cite{Bonin2010OL, hillmann2017off, meleppat2024axis, tateno2026dynamic, Komeda2026BOE} is a variant of OCT that uses a wavelength-sweeping light source and an area camera.
Since FF-SSOCT does not have a confocal pinhole, it is free from confocal signal decay \cite{YueZhu2026OptCon}.
In addition, computational (i.e., holographic) refocusing and aberration correction \cite{Kumar2013OpEx, GZXu2026BOE} enable diffraction-limited resolution over a depth range much wider than the DOF \cite{Kumar2014OpEx, tateno2026dynamic}.
The absence of a confocal pinhole and the use of computational refocusing resolve the resolution-DOF trade-off, and hence, FF-SSOCT can achieve micrometer-scale lateral resolution while maintaining a millimeter-scale imaging depth \cite{YueZhu2026OptCon, tateno2026dynamic, Komeda2026BOE}.

FF-SSOCT enables fast volumetric imaging with high lateral resolution, making it well-suited for 3D imaging of thick \invitro samples, such as tumor spheroids.
However, the coherent nature of OCT inevitably causes speckle \cite{goodman1975statistical,schmitt1999speckle}.
Speckle is a granular pattern that obscures fine sample structures and thus reduces the effective resolution (i.e., the visibility of small structures).
Therefore, effective speckle reduction is essential for microscopic OCT imaging.

There are several speckle reduction methods, but each has its own limitations.
For example, the multi-frame averaging method \cite{gotzinger2011speckle, alonso2011speckle, tan2012speckle} is widely used and highly effective.
However, it requires a large number of OCT images acquired at the same position, resulting in a long acquisition time.
The angle-compounding method \cite{zhao2020angular,desjardins2006speckle} is based on averaging multiple images acquired from several directions.
The wavelength-compounding method \cite{pircher2003speckle} is based on averaging OCT images acquired at different wavelength bands.
The polarization diversity method \cite{storen2004comparison} averages OCT images obtained with different polarization states.
All these methods work well, but they require major hardware modifications.

While the above-mentioned methods are based on modified acquisition protocols or hardware extensions (which we refer to as ``hardware-based methods''), some methods are solely based on image processing (i.e., ``software-based methods'').
Although software-based methods are successful, they still possess some limitations.
For example, digital filter-based methods \cite{adler2004speckle, puvanathasan2007speckle, ozcan2007speckle} can suffer from resolution degradation, expensive computation, or structural distortion.
The non-local-means method effectively suppresses speckle while preserving perceptually important image features, but it is computationally expensive \cite{cuartas2018volumetric, abbasi2024computational}.
Neural network-based methods \cite{ge2024self,chintada2024probabilistic,ni2022hybrid,Ma:18,Yu:23} require large datasets for training.
In addition, both hardware- and software-based methods are often designed without a clear analytical model (i.e., the mathematical definition) of speckle.

Recently, Tomita \etal proposed an analytical signal-formation model of OCT \cite{tomita2023theoretical}.
In this formulation, an OCT image is represented as the sum of two mathematically defined components: an ``incoherent image'' and ``speckle.''
The incoherent image is expressed as a convolution between a real and positive point spread function (PSF) and the sample structure and represents an ideal speckle-free image.
In contrast, speckle arises from interactions among scattered lights from different scatterers and is described as an integral term in this formulation.
Although this speckle term is complicated, it provides a clear mathematical definition of OCT speckle and offers a theoretical basis for manipulating speckle independently from the incoherent image.

In this paper, we exploit this formulation to derive a fully numerical speckle reduction method that does not require specific hardware extensions.
This method is based on model-based numerical complex-signal processing that can modulate speckle patterns after image acquisition while keeping the incoherent image component unchanged.
We generate multiple OCT images with different speckle realizations, and these images are averaged to suppress the speckle.
The imaging model suggests that this operation does not sacrifice OCT resolution.
Since this method uses only a single OCT image, the image acquisition time is short.
In addition, since this method relies only on simple and definitive numerical computation, it requires a short computation time.
We implement this method on an FF-SSOCT system.
FF-SSOCT does not require mechanical lateral scanning, and it possesses high transversal phase stability.
And hence, FF-SSOCT is suitable for our complex-signal-processing-based speckle reduction.
The resolution-preserving nature of this method is experimentally demonstrated by measuring an established scatterer phantom.
This method is then applied to biological samples, including tumor spheroids and a zebrafish eye, and its performance is evaluated subjectively by visual observation and objectively using quantitative image-based metrics.

\section{Principle and theory}
\label{sec:principle}
\subsection{Image-processing flow}
\begin{figure}
	\centering\includegraphics{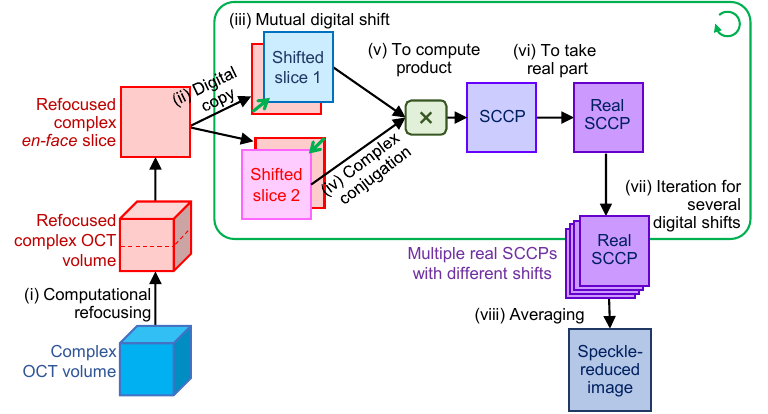}
	\caption{Schematic diagram of the proposed speckle reduction processing.
		The complex OCT image is refocused (i), duplicated (ii), and mutually shifted (iii).
		One of the duplicated images is complex-conjugated (iv) and multiplied by the other image (v) to yield a ``shifted-complex-conjugate product (SCCP)'' image.
		Subsequently, the real part of the SCCP (real SCCP) is computed (vi).
		This process is iterated for several shift directions and amounts (vii), resulting in real SCCP images that have different speckle realizations, as detailed in Section \ref{sec:mathematics}.
		These real SCCP images are finally averaged (viii) to obtain the speckle-reduced image.
	}
	\label{fig:SpeckleReductionSchematic}
\end{figure}
Figure \ref{fig:SpeckleReductionSchematic} illustrates the overall processing flow of the proposed speckle reduction method.
Prior to speckle reduction, we first correct the defocus of the complex OCT image by computational refocusing [Step-(i) of Fig.\@ \ref{fig:SpeckleReductionSchematic}].
The details of this computational refocusing method can be found elsewhere \cite{tateno2026dynamic}.

After computational refocusing, an \enface complex OCT image is extracted from the volume.
This image is duplicated (Step-(ii)), and the copies are mutually digitally shifted (Step-(iii)).
The product of one of the images and the complex conjugate of the other image is computed, as depicted in Steps-(iv) and (v).
Here, we denote this product as the ``shifted-complex-conjugate product (SCCP).''
Subsequently, the real part of the SCCP (denoted as ``real SCCP'') is computed (Step-(vi)).
As will be shown in the next section (Section \ref{sec:mathematics}), the real SCCP consists of two components: an incoherent image of the sample and speckle-related components.
For simplicity, we denote this speckle-related component as speckle.

This process is iterated for several shift directions and amounts (Step-(vii)).
It results in multiple real SCCP images with the same incoherent image component but with different speckle realizations (see the next section for theoretical details).
These real SCCP images are finally averaged (Step-(viii)), and the speckle-reduced image is obtained.

For image observation, the speckle-reduced image is displayed on a logarithmic (dB) scale after taking its absolute value.
It is noteworthy that both the standard OCT intensity and the real SCCP are products of a complex OCT signal and a complex-conjugate OCT signal.
Hence, they have the same signal dimension, which makes the dB-scale display reasonable.

\subsection{Theory}
\label{sec:mathematics}
\subsubsection{Model of complex OCT signal}
The theory of our speckle reduction method starts from the OCT image formation model presented by Tomita \etal \cite{tomita2023theoretical}.
In this model, the sample to be measured is represented as a spatially slowly varying refractive index distribution in which scatterers are spatially randomly dispersed.
With this sample representation, the complex OCT signal at a particular depth in the OCT volume is expressed as follows
\begin{equation}
	\label{eq:complexOct}
	s(\bmxp) = \sum_{i=1}^{N} B(\bmx_i) e^{i\phi_i} P_c(\bmxp - \bmx_i),
\end{equation}
where $\bmxp = (x', y')$ is the two-dimensional lateral position in the image, and $\bmx_i = (x_i, y_i)$ is the lateral position of the $i$-th scatterer in the sample.
Here, the magnification between the sample and the image is assumed to be 1 without loss of generality.
$N$ is the total number of scatterers in the sample, $B(\bmx_i)$ is the scattering potential distribution of the sample, and $P_c$ is the complex PSF of the OCT system.
$\phi_i$ is the phase offset caused by the depth position of the $i$-th scatterer, and it is practically unpredictable and random.
In total, the complex OCT signal is the weighted summation of scattered light from each scatterer, with the PSF serving as the weighting function.
Note that our method is based on \enface image manipulation, and hence, we focus on the \enface formulation here.

The original formulation by Tomita \etal is for point-scanning OCT, but the complex PSF of FF-SSOCT is not identical to that of point-scanning OCT.
For point-scanning OCT, the PSF is defined by both the illumination and collection optics.
On the other hand, the PSF of FF-SSOCT is governed only by the collection optics because the sample is illuminated by a flood plane wave (namely, the pupil function of the illumination is a delta function).
As shown in Refs.\@ \cite{YueZhu2026OptCon, tateno2026dynamic} in detail, the lateral PSF of FF-SSOCT is
\begin{equation}
	\label{eq:GaussianPSF}
	P_c(\bmx, z;z_0) = \frac{w_0}{w(z;z_0)}
	\exp\left[{- \frac{\bmx^2}{w^2(z;z_0)}}\right]
	\exp \left[-i \left\{n k_0 (z-z_0) + \frac{n k_0 \bmx^2}{2R(z;z_0)} - \psi(z;z_0) \right\}\right],
\end{equation}
where $z_0$ is the depth position of the focus,
$w(z;z_0)$ is the radius of the probe beam at depth $z$,
$w_0 = w(z = z_0; z_0)$ is the collection beam radius (i.e., a virtual beam spot radius on a sample assuming a point source is at the camera plane) at the in-focus depth.
$n$ is the refractive index of the sample around the position of interest, and $k_0$ is the center wavenumber of the light source.
The second term in the second exponential represents the quadratic phase induced by defocus, and $R(z;z_0)$ is the phase curvature induced by the defocus.
$\bmx^2$ represents $\bmx \cdot \bmx$.
Similarly, hereafter, a squared vector represents the inner product of a vector with itself.
In the third term of the second exponential, $\psi(z;z_0)$ represents the Gouy phase.
The defocus-induced phase curvature $R$ is given by
\begin{equation}
    R(z;z_0)= \rbra{z-z_0} \abra{1 + \cbra{\frac{n^2 \, k_0 \, w_0^2}{2 (z-z_0)}}^2 }.
	\label{eq:R}
\end{equation}

By specifying the depth position $z$, Eq.\@ (\ref{eq:GaussianPSF}) can be substituted into Eq.\@ (\ref{eq:complexOct}).

\subsubsection{Shifted-complex-conjugate product (SCCP) and real SCCP}
Following the same mathematical flow as Ref.\@ \cite{tomita2023theoretical} but with the complex PSF of Eq.\@ (\ref{eq:GaussianPSF}), the analytic representation of the SCCP of FF-SSOCT is derived as
\begin{equation}
	\begin{split}
		& \mathrm{SCCP} = s\!\left(\bmxp - \frac{\deltax}{2}\right)
		s^*\!\left(\bmxp + \frac{\deltax}{2}\right)
		={}\,
		{\color{blue}
			\frac{w_0^2}{w^2}
			\exp\!\left(-\frac{\deltax^2}{2w^2}\right)
			\int_{-\infty}^{\infty} D(\bmx)\, B^2(\bmx)
		} \\
		&\qquad\qquad
		{\color{blue}
			\times \exp\!\left[-\frac{2}{w^2}(\bmxp - \bmx)^2\right]
			\exp\!i\!\left[\frac{n k_0}{R}\,\deltax\cdot(\bmxp-\bmx)\right]\,\dd\bmx
		}\\
		&\quad\qquad +\,
		{\color{red}
			\frac{w_0^2}{w^2}
			\exp\!\left(-\frac{\deltax^2}{2w^2}\right)
			\iint_{-\infty}^{\infty} D'(\bmxpp,\bmxqq)\,B(\bmxpp)B(\bmxqq)
		}\\
		&\qquad\qquad
		{\color{red}
			\times \exp\!\left[-\frac{1}{w^2}(\bmxpp-\bmxqq)\!\cdot\!\left(\frac{\bmxpp-\bmxqq}{2}+\deltax\right)\right]
			e^{i\phi_{pq}} \exp\!\left[-\frac{2}{w^2}\!\left(\bmxp-\frac{\bmxpp+\bmxqq}{2}\right)^{\!2}\right]
		}\\
		&\qquad\qquad
		{\color{red}
			\times \exp\!i\!\left[\frac{n k_0}{R}(\bmxpp-\bmxqq+\deltax)\!\cdot\!\left(\bmxp-\frac{\bmxpp+\bmxqq}{2}\right)\right]
			\,\dd\bmxpp\,\dd\bmxqq
		}\\
		&\quad\qquad -\ \text{the first integral of R.H.S.,}
	\end{split}
	\label{eq:SCCP}
\end{equation}
where R.H.S.\@ means ``the right-hand side.''
Here, $D$ is the scatterer density distribution, and $D'$ is the cross-density of the scatterers, which is a spatial function closely related to $D$.
$\phi_{pq}$ is a practically random phase due to the depth positions of scatterers $p$ and $q$.
See Ref.\@ \cite{tomita2023theoretical} for more details about these quantities.
$\deltax$ is the lateral two-dimensional shift digitally introduced after data acquisition (Step-(iii) in Fig.\@ \ref{fig:SpeckleReductionSchematic}).
Notably, when the shift is zero, the SCCP becomes identical to the conventional OCT intensity image.
Here, the blue parts of the equation correspond to the incoherent image component, while the red parts correspond to speckle.

The real SCCP is obtained by taking the real part of this SCCP as
\begin{equation}
	\begin{split}
		& \mathrm{Re}\left[
		s\!\left(\bmxp - \frac{\deltax}{2}\right)
		s^*\!\left(\bmxp + \frac{\deltax}{2}\right)
		\right]
		={}\,
		{\color{blue}
			\frac{w_0^2}{w^2}
			\exp\!\left(-\frac{\deltax^2}{2w^2}\right)
			\int_{-\infty}^{\infty} D(\bmx)\, B^2(\bmx)
		}\\
		&\quad\qquad
		{\color{blue}
			\times \exp\!\left[-\frac{2}{w^2}(\bmxp - \bmx)^2\right]
			\cos\!\left[\frac{n k_0}{R}\deltax\cdot(\bmxp - \bmx)\right]
			\, \dd\bmx
		}\\
		&\qquad +\,
		{\color{red}
			\frac{w_0^2}{w^2}
			\exp\!\left(-\frac{\deltax^2}{2w^2}\right)
			\iint_{-\infty}^{\infty} D'(\bmxpp, \bmxqq)\, B(\bmxpp)B(\bmxqq)
		}\\
		&\qquad\quad
		{\color{red}
			\times \exp\!\left[-\frac{1}{w^2}(\bmxpp - \bmxqq)
			\!\cdot\!\left(\frac{\bmxpp - \bmxqq}{2} + \deltax\right)\right]
			\exp\!\left[-\frac{2}{w^2}\!\left(\bmxp - \frac{\bmxpp + \bmxqq}{2}\right)^{\!2}\right]
		}\\
		&\qquad\quad
		{\color{red}
			\times \cos\!\left[\phi_{pq} + \frac{n k_0}{R}\,(\bmxpp - \bmxqq + \deltax)
			\!\cdot\!\left(x' - \frac{\bmxpp + \bmxqq}{2}\right)\right]
			\, \dd\bmxpp\, \dd\bmxqq
		}\\
		&
		\qquad{}-{}\ \text{the first integral of R.H.S.}
	\end{split}
	\label{eq:Re[SCCP]}
\end{equation}
Here, we used the fact that $D$ (the scatterer density distribution), $D'$ (the cross-density distribution of scatterers), and $B^2$ (the squared scattering potential) take real and positive values.
The only complex components of Eq.\@ (\ref{eq:SCCP}) are the exponential functions containing the imaginary unit ($i$), which became cosine functions according to Euler's formula in Eq.\@ (\ref{eq:Re[SCCP]}).

\subsubsection{Speckle modulation and reduction}
By computationally correcting the defocus (Step-(i) of Fig.\@ \ref{fig:SpeckleReductionSchematic}), $w$ in Eq.\@ (\ref{eq:Re[SCCP]}) becomes $w_0$ (virtual in-focus collection beam radius), and the phase curvature $R$ approaches infinity.
Thus, the real SCCP is simplified to
\begin{equation}
	\begin{array}{l}
		\mathrm{Re}\!\left[
		s\!\left(\bmxp - \dfrac{\deltax}{2}\right)
		s^*\!\left(\bmxp + \dfrac{\deltax}{2}\right)
		\right]
		= \\
		\qquad\qquad{\color{blue}
			\displaystyle
			\exp\!\left(-\frac{\deltax^2}{2w_0^2}\right)
			\int_{-\infty}^{\infty} D(\bmx) B^2(\bmx)
			\exp\!\left[-\frac{2}{w_0^2}(\bmxp - \bmx)^2\right]
			\dd\bmx
		} \\
		\qquad\qquad
		{}+{} {\color{red}
			\displaystyle
			\exp\!\left(-\frac{\deltax^2}{2w_0^2}\right)
			\iint_{-\infty}^{\infty} D'(\bmxpp, \bmxqq)
			B(\bmxpp) B(\bmxqq)
			\exp\!\left[-\frac{1}{w_0^2}(\bmxpp - \bmxqq)
			\cdot \left(\frac{\bmxpp - \bmxqq}{2} + \deltax\right)\right]
		} \\
		\qquad\qquad\quad {\color{red}
			\displaystyle
			{}\times{}
			\exp\!\left[-\frac{2}{w_0^2}\left(\bmxp - \frac{\bmxpp + \bmxqq}{2}\right)^2\right]
			\cos(\phi_{pq})\, \dd\bmxpp\, \dd\bmxqq
		} \\
		\qquad\qquad{}-{} \text{the first integral of R.H.S.}
	\end{array}
	\label{eq:Re[SCCP]_nodefocus}
\end{equation}
As we have mentioned, the blue part of the equation corresponds to the incoherent image.
Namely, the integral represents the convolution of $D(\bmx) B^2(\bmx)$ and $\exp\left[-2(\bmx)/w_0^2\right]$.
The former corresponds to the scattering-strength distribution of the sample, which can be considered the ``sample structure'' measured by OCT.
On the other hand, the latter is a real positive Gaussian function having the same size as the lateral intensity PSF of the OCT system.
Since both of them are real and positive, this blue part corresponds to the incoherent image of the sample.
Note that this blue part is insensitive to the digital shift ($\deltax$) except for the overall magnitude-scaling factor, i.e., the exponential part to the left of the first integral.
Hence, this blue part is analogous for any shift values.

On the other hand, the red part corresponding to speckle contains $\deltax$ within the integral and is therefore modulated by changing $\deltax$.

In summary, Eq.\@ (\ref{eq:Re[SCCP]_nodefocus}) implies that the speckle pattern of the real-SCCP image can be altered by changing the digital shift even after the OCT acquisition (i.e., physical measurement), while the same operation keeps the incoherent image component unchanged.
And hence, each of the multiple real SCCPs in Fig.\@ \ref{fig:SpeckleReductionSchematic} has different speckle realizations.
Finally, speckle can be reduced by averaging the multiple real-SCCP images (Step-(viii) of Fig.\@ \ref{fig:SpeckleReductionSchematic}).

\section{Implementation and study protocol}
\subsection{FF-SSOCT system}
A custom-built FF-SSOCT system \cite{tateno2026dynamic} was used to validate our speckle reduction method.
The light source is a wavelength-sweeping light source with a center wavelength of 840 nm and a sweeping range of 75 nm (BS-840-1-HP, Superlum, Ireland).
The probe beam illuminates the sample as a collimated plane wave, where the beam diameter and the power on the sample are 2.3 mm and 5.2 mW, respectively.
\memo{2.34 mm and 5.23 mW.}
The sample is imaged on a high-speed area camera (FASTCAM Mini AX100, Photron, Japan) through a microscope objective (NA = 0.3, effective focal length = 18 mm, RMS10X-PF, Olympus, Japan) and a tube lens ($f$ = 750 mm).
The backscattered probe beam overlaps with a reference beam and forms an interference signal on the camera.

The in-focus lateral resolution at full width at half maximum (FWHM) intensity is 1.4 \um (in air), and the axial resolution is 6.5 \um (FWHM intensity in air).
The camera pixel size is 20 \um, which corresponds to 0.48 \um on the sample and is approximately 1/3 of the in-focus lateral resolution.
The DOF is 16.0 \um.
The sensor resolution is $1024 \times 1024$ pixels, which results in a maximum lateral field of view (FOV) of 0.49 mm $\times$ 0.49 mm.
The camera operates at 4,000 frames/s, and a single OCT volume is reconstructed from 1,000 spectral interference signals.
Hence, the volumetric measurement speed is 4 volumes/s.

After the OCT volume is reconstructed, the defocus is computationally corrected throughout the entire imaging depth using a phase-only spatial-frequency filter.

More details of the OCT system and the computational refocusing are described elsewhere \cite{tateno2026dynamic}.

\subsection{Implementation of the speckle reduction method}
As described in Principle (Section \ref{sec:principle}), the speckle reduction method generates multiple real SCCP images with several shifts (i.e., several shift amounts and directions) and averages them.
These shifts are defined as follows.

\begin{figure}
	\centering
	\includegraphics[width=\linewidth]{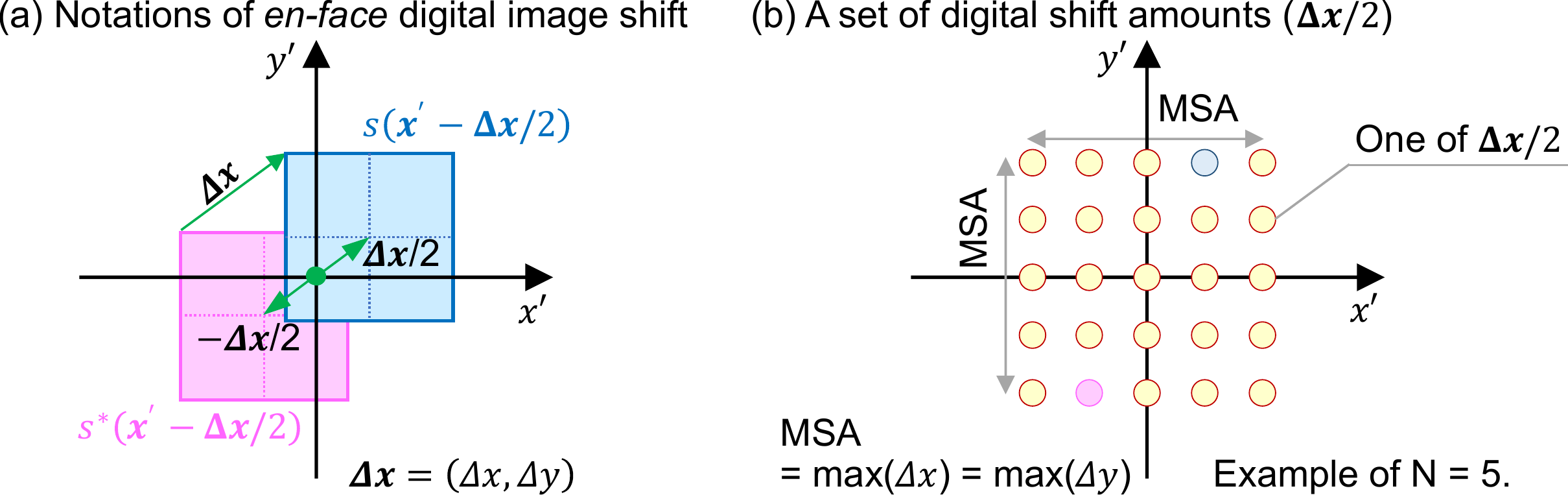}
	\caption{Diagrams representing the set of digital shifts.
		(a) illustrates the spatial relationship between two shifted images and the original image.
		(b) illustrates the set of shifts for one image. Each point in this diagram represents one shift pattern.
		The shifts of the other image are point-symmetric to these points around the origin.
	}
	\label{fig:shiftsSchematic}
\end{figure}

Figure \ref{fig:shiftsSchematic}(a) illustrates an example of the digital shift operation.
Here, the complex \enface image $s$ of Eqs.\@ (\ref{eq:SCCP})--(\ref{eq:Re[SCCP]_nodefocus}) is digitally shifted by $\deltax/2$, and $s^*$ is shifted by $-\deltax/2$.
The mutual displacement of these two images is $\deltax = (\Delta x, \Delta y)$.

In our speckle reduction method, multiple real-SCCP images with several shifts are generated and averaged.
A set of shifts is exemplified in Fig.\@ \ref{fig:shiftsSchematic}(b), where each circle indicates one of the shifts of the $s$ image, i.e., $\deltax/2$.
The shifts in a set are selected to form a rectangular equi-spacing  grid, as shown in this diagram.
The numbers of shift patterns along the $x'$ and $y'$ directions are identical and denoted as $N$.
In this example, $N = 5$, and there are 25 (= $N^2$) shift patterns in total.
It should be noted that when a complex image $s$ shifts by $\deltax / 2$, the other image $s^*$ shifts by $-\deltax /2$.
Namely, when the shift of $s$ is the blue circle in the figure, the shift of $s^*$ is the purple circle.

We define the parameter ``maximum shift amount'' (MSA) as the maximum of $\left|\Delta x\right|$, which is identical to the maximum of $\left|\Delta y\right|$ in our configuration, as depicted in Fig.\@ \ref{fig:shiftsSchematic}(b).

Each shift (the circle in Fig.\@ \ref{fig:shiftsSchematic}(b)) corresponds to a single real SCCP image, and all of the real-SCCP images are averaged to reduce speckle.
It should be noted that some shifts yield identical real-SCCP images.
For example, the SCCP obtained with the shift of the blue circle in Fig.\@ \ref{fig:shiftsSchematic}(b) is the complex conjugate of the SCCP obtained with the purple circle.
Hence, the real-part images (i.e., real-SCCP images) of these SCCPs are identical.
This suggests that ``the number of unique images'' (NUI) to be averaged is not $N^2$ but $(N^2 +1)/2$.

In some shift configurations, the digital shift amount is not always an integer multiple of the pixel size.
For shifts with sub-pixel accuracy, the complex \enface OCT image is up-sampled using a Fourier-space zero-padding method in our specific implementation.

\subsection{Sample and performance-evaluation method}
\subsubsection{Samples}
\label{subsec:Samples}
To quantitatively evaluate the impact of our speckle reduction method on lateral resolution, a sparse-scatterer phantom (OCT point spread function phantom, OCTPSF01, Edmund Optics, NJ) was measured.
The phantom consists of a solid medium with a refractive index of approximately 1.5 and a thickness of 10 mm, in which FeO nanoparticles smaller than 1 \um are sparsely suspended.
The randomly distributed scatterers generate isolated bright spots in OCT images, enabling quantitative assessment of the lateral resolution.

To validate the speckle-reduction effect of the present method in biomedical samples, \invitro cancer spheroids and a postmortem fixed adult zebrafish (AB strain) were measured.
The spheroids were formed by seeding 1,000 MCF-7 (human breast adenocarcinoma cell line) cells.
To include a variety of samples, we used three spheroids with different cultivation protocols.
Namely, two of them were cultured without anti-cancer drug administration, where one was cultured for 8 days and the other was cultured for 10 days.
The remaining one was cultured for 8 days with 10 \uM of doxorubicin (DOX, an anticancer drug).

The zebrafish sample is an 11-month-old wild-type zebrafish.
The fish were euthanized by administering an overdose of tricaine, fixed in a 4\% paraformaldehyde solution, and stored in phosphate-buffered saline (PBS).
All experimental protocols involving zebrafish were approved by the Institutional Animal Care and Use Committee (IACUC) of the University of Tsukuba.

\subsection{Image evaluation metrics}
To quantitatively evaluate speckle-reduction performance, two widely used image metrics are employed: the contrast-to-noise ratio (CNR) and the equivalent number of looks (ENL).

The CNR represents the contrast of sample structures in the image and is defined as
\begin{equation}
	\mathrm{CNR} = 10 \log_{10} \left(
	\frac{ \left| \mu_1 - \mu_2 \right| }
	{ \sqrt{ \sigma_1^2 + \sigma_2^2 } }
	\right),
	\label{eq:CNR}
\end{equation}
where $\mu_1$ and $\mu_2$ represent the mean values of the image in the high-intensity region (region of interest 1, ROI-1) and the low-intensity region (ROI-2), respectively, and $\sigma_1^2$ and $\sigma_2^2$ are the variances within ROIs-1 and -2, respectively.

The ENL represents the smoothness of the image within a homogeneous region and is defined as
\begin{equation}
	\mathrm{ENL} = \frac{ \mu^2 }{ \sigma^2 },
	\label{eq:ENL}
\end{equation}
where $\mu$ denotes the mean signal within the selected ROI and $\sigma^2$ denotes the corresponding variance.

\section{Results}
\label{sec:Result}
\subsection{Resolution analysis by scattering phantom}
\label{subsec:Resolution preservation analysis}
\begin{figure}
	\centering
	\includegraphics[width=\linewidth]{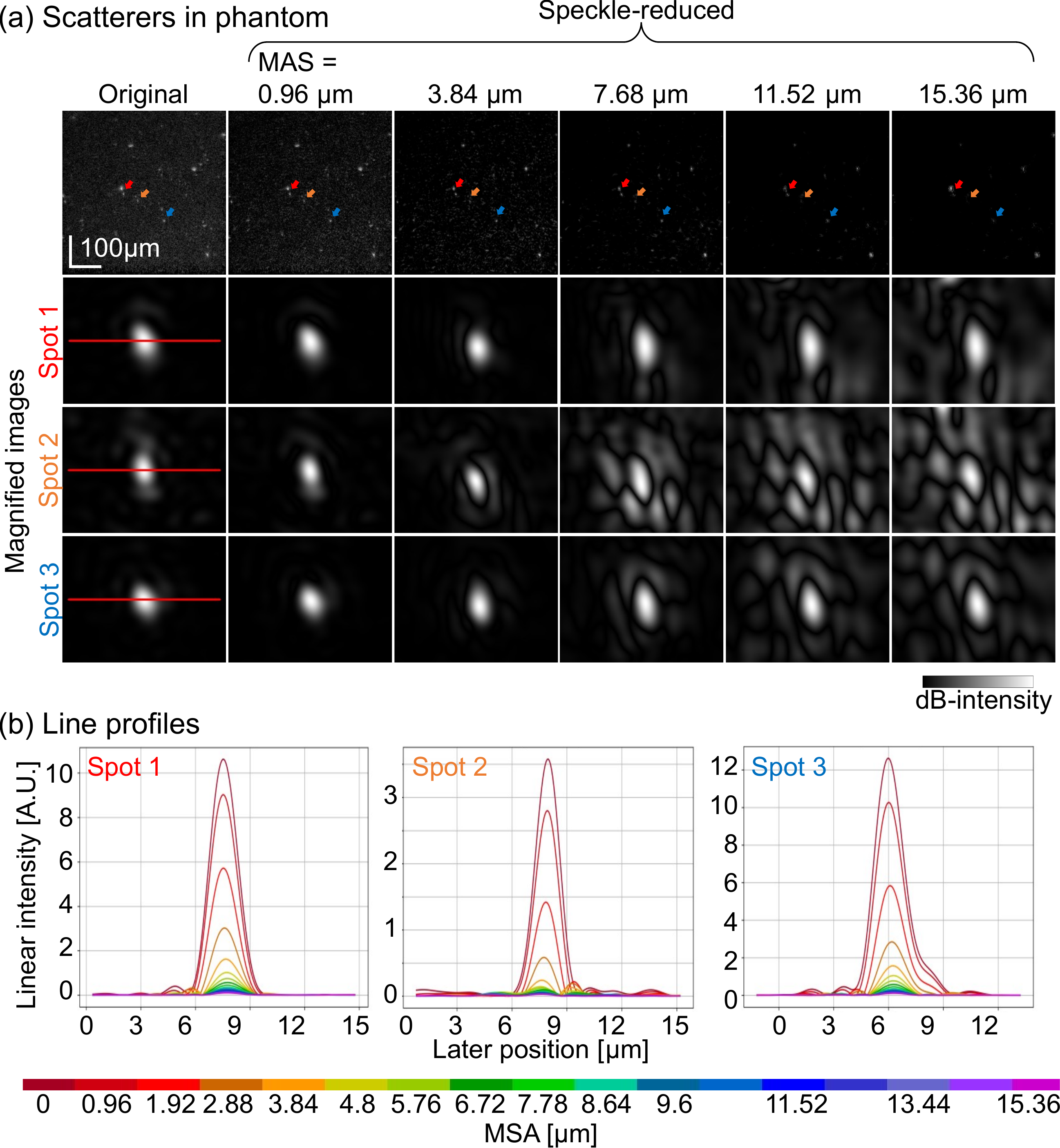}
	\caption{
		Resolution evaluation using a sparse-scatterer phantom.
		(a) \Enface OCT image of the phantom (column 1) and its speckle-reduced images generated with an NUI of 145 (N=17) and several MSAs (columns 2-6).
		The magnified images of three representative scattering spots are displayed in rows 2-4.
		Images are shown on a dB scale, and the magnified images are normalized to have the same peak intensity.
		(b) Line profiles of the spots in the original and speckle-reduced images, where the plot color represents the MSA.
		The profiles are obtained along the red lines in (a).
		The spot width, which can be considered as the lateral resolution, is insensitive to the MSA, while the peak intensity of the spot decreases as the MSA increases.
	}
	\label{fig:psf_phantom}
\end{figure}
Figure \ref{fig:psf_phantom}(a) shows the \enface OCT image of the sparse-scatterer phantom (column 1) and the speckle-reduced images obtained with $N = 17$ (NUI = 145) and several MSAs (columns 2-6).
It can be seen that the proposed speckle-reduction method maintains the sharpness of the isolated scattering spots.

\begin{table}
	\centering
	\caption{Averaged FWHM of three selected spots in phantom images under different maximum shift amounts.}
	\label{tab:fwhm_shift}
	\begin{tabular}{c|c}
		\hline
		MSA & FWHM of spot (mean $\pm$ standard deviation) \\
		\hline
		0.00 \um & 1.76 $\pm$ 0.101 \um \\
		0.96 \um & 1.72 $\pm$ 0.113 \um \\
		3.84 \um & 1.52 $\pm$ 0.185 \um \\
		7.68 \um & 1.56 $\pm$ 0.098 \um \\
		11.52 \um & 1.60 $\pm$ 0.057 \um \\
		15.36 \um & 1.64 $\pm$ 0.075 \um \\
		\hline
	\end{tabular}
\end{table}
Three representative spots (spots 1 to 3) are enlarged in rows 2 to 4 in the figure.
The enlarged images are shown on a normalized dB scale in which the main peaks appear with the same brightness across the images.
The line profiles of these spots along the red lines are shown in Fig.\@ \ref{fig:psf_phantom}(b), where the profiles are plotted on a linear-intensity scale and the plot color represents MSA.
The mean and standard deviation of the FWHMs of the three spots are summarized in Table \ref{tab:fwhm_shift} for each MSA.
It can be seen that the sharpness of the main peak is preserved even with a large MSA [Table \ref{tab:fwhm_shift} and the line profiles].
However, the peak height is reduced as the MSA increases as shown in the line profiles, and it is expected due to the $\deltax$-dependent intensity scaling factor that appears in Eq.\@ (\ref{eq:Re[SCCP]}) (the first part of the right-hand side: $w_0^2/w^2 \exp [-\deltax^2/(2w^2)]$), as shown in the line profiles.
Consequently, in some spots (such as spots 2 and 3), originally low-intensity artifactual peaks around the main peak are pronounced in the normalized-dB-scaled enlarged image.
Nevertheless, it is still worth reemphasizing that despite the pronunciation of these artifacts in some cases, the main peaks retain their original sharpness in all cases.
Furthermore, these artifacts do not cause significant disturbance in practical cases, as shown later with biological samples.

\subsection{Tumor spheroids}
\begin{figure}
	\centering\includegraphics[width=4.87in]{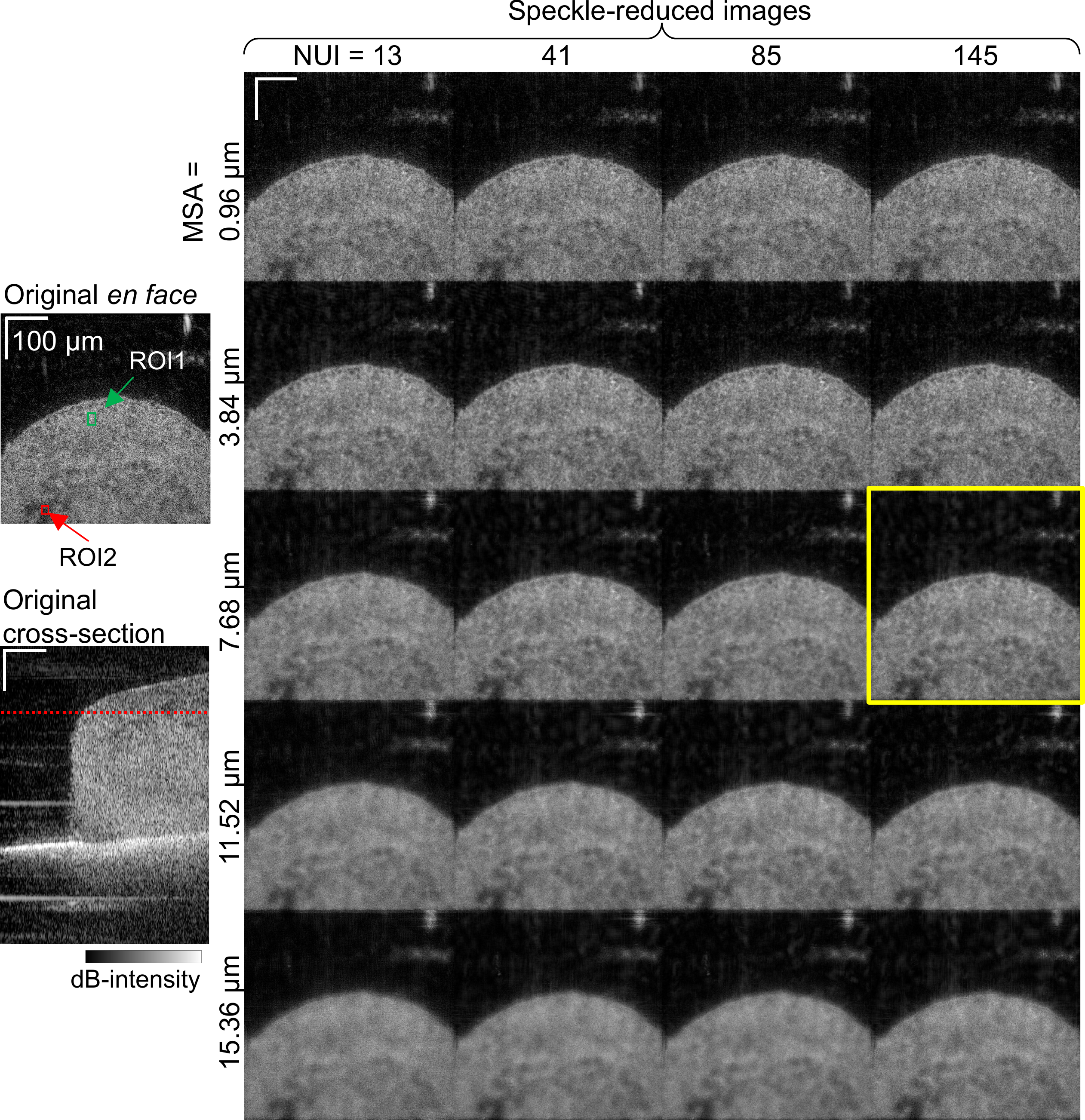}
	\caption{Original and speckle-reduced images of a cancer spheroid.
		The left column displays the original \enface image and its corresponding cross-sectional image, where the red dashed line indicates the depth of the \enface slice.
		The grid of images on the right represents speckle-reduced images generated with various combinations of MSA and NUI.
		The scale bars represent 100 \um $\times$ 100 \um.
	}
	\label{fig:ParameterInvestigation}
\end{figure}
Figure \ref{fig:ParameterInvestigation} illustrates speckle-reduced images of a tumor spheroid generated using various MSA and NUI configurations.
According to visual observation, a larger MSA results in a greater reduction in the appearance of speckle.
However, when the MSA becomes too large, the image becomes blurry and the structural details of the sample become invisible.
This could be partially accounted for by the artifactual side peaks associated with a large MSA, as discussed in Section \ref{subsec:Resolution preservation analysis}.
Uncorrected high-order wavefront aberrations and residual defocus can also cause this blurring, since our theory assumes that the images do not suffer from defocus and aberrations.

It is also found that variations in NUI do not produce a substantial difference in image quality, although the case with NUI = 145 shows slightly better performance than the others.
In addition, increasing the NUI does not significantly increase the computational load.
Therefore, $(\mathrm{MSA}, \mathrm{NUI}) = (7.68\,\mu\mathrm{m}, 145)$, which is indicated by a yellow box in the figure, can be a reasonable choice for practical use.

\begin{figure}
	\centering\includegraphics{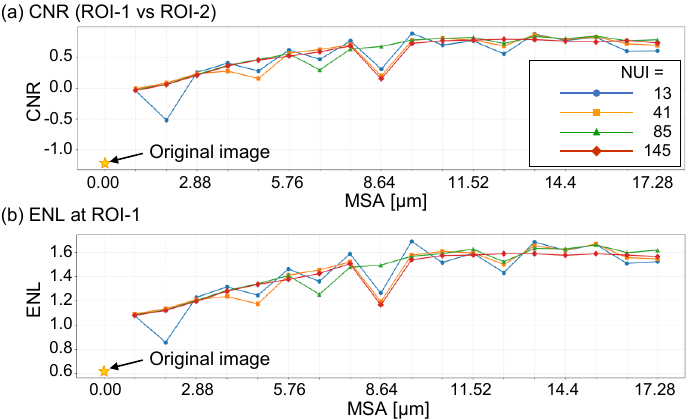}
	\caption{
		CNR (a) and ENL (b) with several (MSA, NUI) configurations.
		The CNR and ENL represent the image contrast and the smoothness of the image, respectively.
		Both CNR and ENL increase (i.e., improve) as MSA increases.
		The CNR and ENL are computed using two ROIs indicated in the top-right image of Fig.\@ \ref{fig:ParameterInvestigation}.
	}
	\label{fig:ImageMetricResults}
\end{figure}
To quantitatively investigate image quality, CNR and ENL are computed from two ROIs (indicated in the top-left image of Fig.\@ \ref{fig:ParameterInvestigation}), where ROI-1 and ROI-2 are high- and low-intensity ROIs, respectively.
The computed CNR and ENL for each NUI are plotted as functions of MSA in Fig.\@ \ref{fig:ImageMetricResults}.
In comparison to the original image (star mark), all configurations of MSA and NUI exhibit higher image contrast (higher CNR) and a smoother image appearance (higher ENL).
Both CNR and ENL become higher (better) for larger MSA values, while the NUI does not introduce a significant difference in CNR and ENL, at least for $\mathrm{NUI} \geq 13$.

By considering both subjective visual observation (Fig.\@ \ref{fig:ParameterInvestigation}) and the CNR-and-ENL analysis (Fig.\@ \ref{fig:ImageMetricResults}), we can conclude that a parameter set of (MSA, NUI) = (7.68 \um, 145) is reasonable.

\begin{figure}
	\centering\includegraphics[width=4.84in]{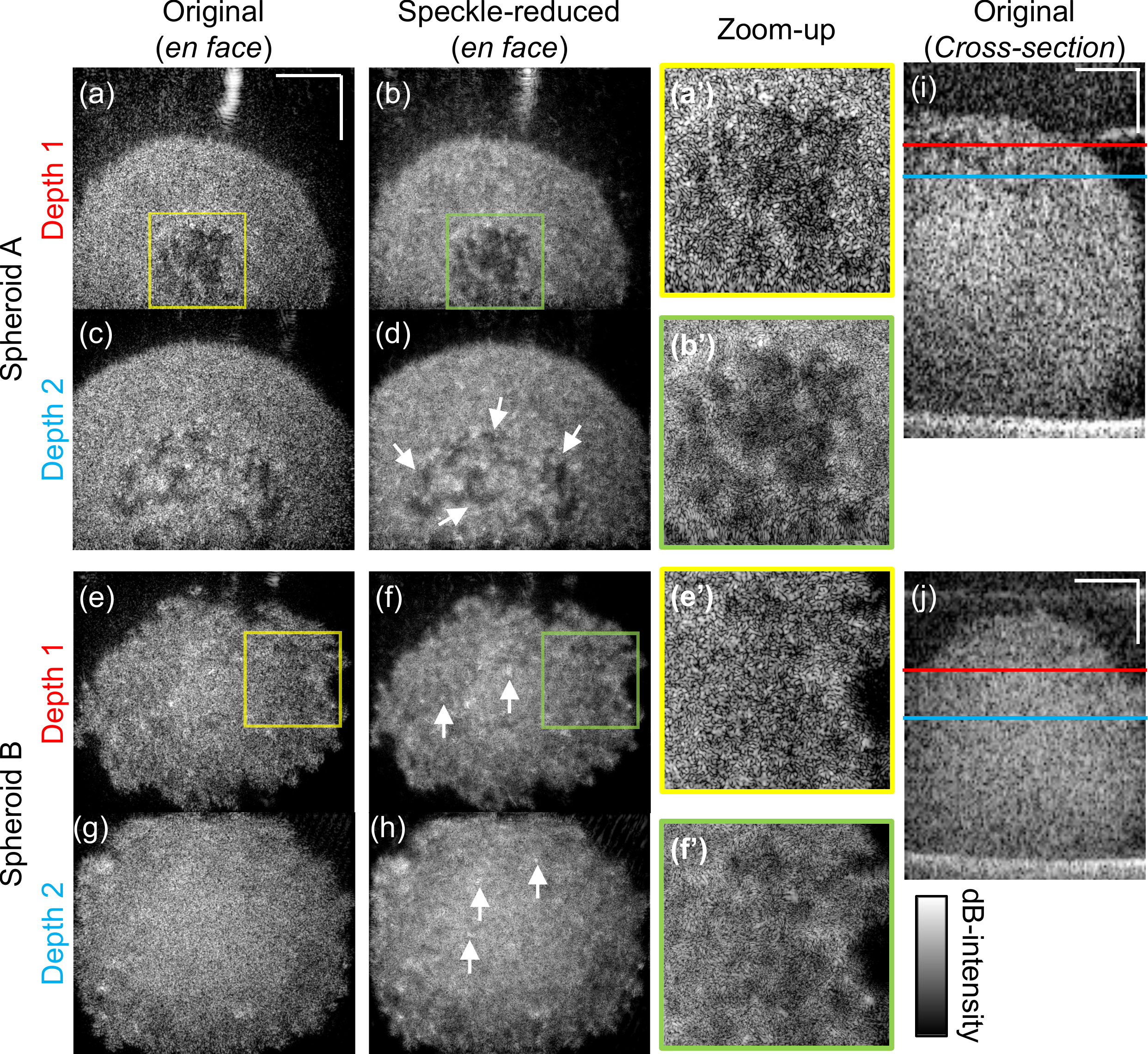}
	\caption{
		Examples of speckle reduction in tumor spheroids.
		Spheroid A was cultivated for 8 days without drug administration, while spheroid B was treated with an anti-cancer drug (doxorubicin) for 3 days after the 8-day cultivation.
		The real-SCCP-based speckle reduction was applied with an MSA of 7.68 \um and an NUI of 145.
		The speckle reduction improved the visibility of structural defects (red arrows in d) and hyper-scattering spots (red arrows in h).
		The scale bars indicate 100 \um, and the depths of the \enface images are indicated by colored dashed lines in the cross-sectional images (i, j).
	}
	\label{fig:OptimalSpheroid}
\end{figure}
Figure \ref{fig:OptimalSpheroid} shows examples of two spheroids under different drug-administration conditions.
The speckle-reduced images (column 2) were generated with an MSA of 7.68 \um and an NUI of 145.
In spheroid A, there are several defects in the core region (exemplified by arrows), which may be caused by necrosis.
The visibility of the structural details of these defects is significantly improved by the speckle reduction.
In spheroid B, the speckle reduction revealed several high-intensity spots (exemplified by arrows).

\subsection{Zebrafish eye}
\begin{figure}
	\centering\includegraphics[width=4.87in]{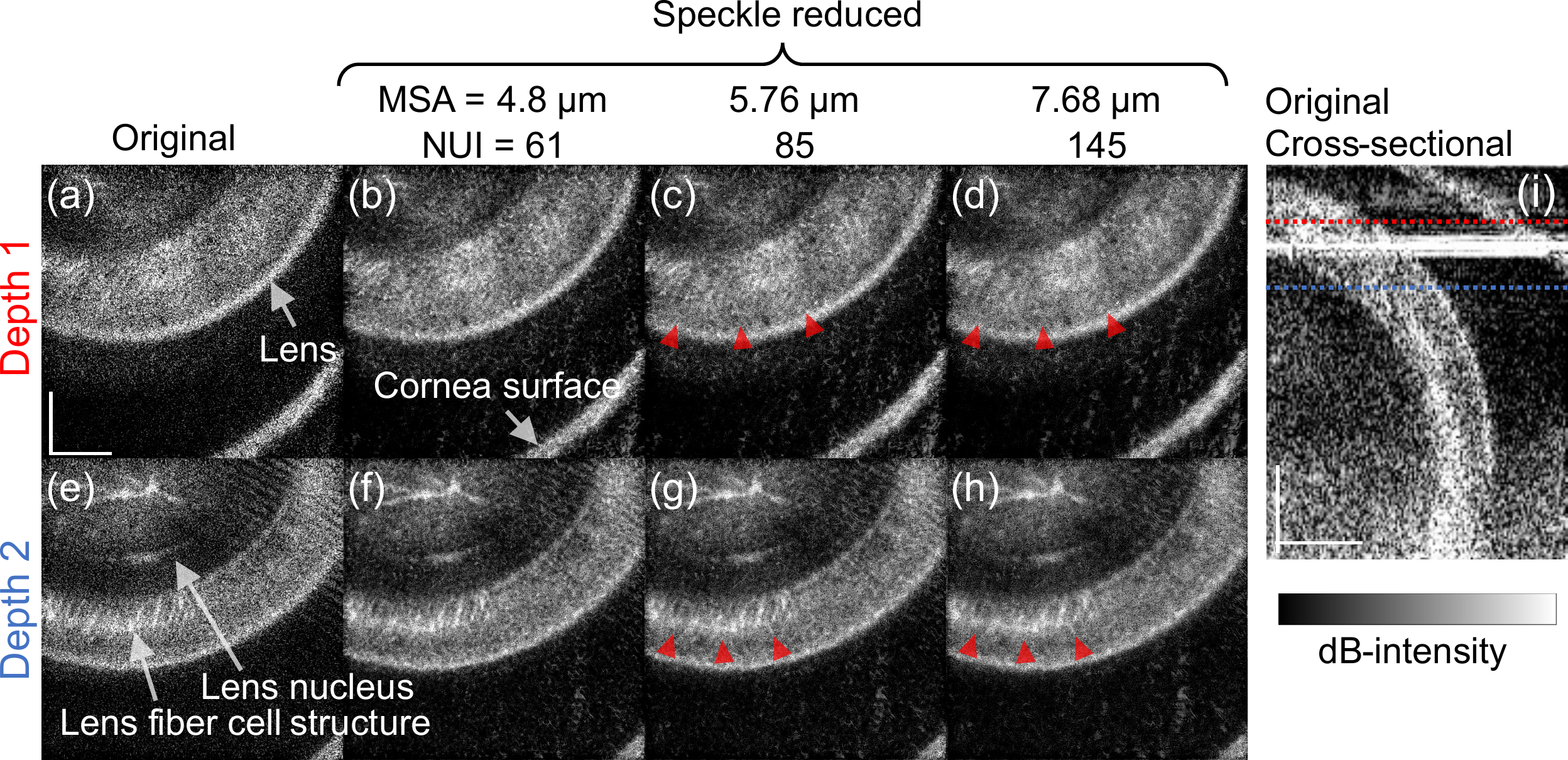}
	\caption{
		Speckle-reduction example of a zebrafish eye.
		The speckle-reduced images (b-d) and (f-h) were generated using the MSA and NUI indicated above the images.
		According to subjective observation, the speckle-reduced images of (c) and (g) provided the best performance.
		The visualization of structure corresponding to lens fiber cells demonstrates the resolution-preserving nature of the real-SCCP-based speckle reduction method.
		The scale bars indicate 100 \um.
	}
	\label{fig:zebrafish}
\end{figure}

Figure \ref{fig:zebrafish} demonstrates speckle reduction in a zebrafish eye.
The speckle-reduced images show clearer microstructures, especially in Figs.\@ \ref{fig:zebrafish}(c) and (g), which are processed with $(\mathrm{MSA}, \mathrm{NUI}) = (5.76\,\mu\mathrm{m}, 85)$.
Specifically, the structure corresponding to the lens fiber cells are clearly visible in Figs.\@ \ref{fig:zebrafish}(c) and (g).
This indicates the resolution-preserving nature of the real-SCCP-based speckle reduction.

It should be noted that an MSA of 7.68 \um, which was the optimal MSA for the spheroid, results in a slightly blurry image appearance in the zebrafish eye.
This may suggest that the optimal MSA depends on the structural characteristics of the sample.

\section{Discussion}
\subsection{Computation time}
To evaluate the computational overhead of the proposed method, we measured both the isolated execution time of the speckle reduction algorithm and the overall post-processing time.
Benchmarking was performed on a standard $1024 \times 1024 \times 512$ pixel OCT volume using a Jupyter-based Python environment.

The core signal processing of the real-SCCP speckle reduction includes digital shifting, complex-conjugate product computation, real-part extraction, repeating these steps for multiple shift patterns, and averaging the real-SCCP images (Step-(ii) to Step-(vii) in Fig.\@ \ref{fig:SpeckleReductionSchematic}).
This core process took approximately 0.55 s per \enfaceh frame with an NUI of 145 ($N = 17$), where sub-pixel shifts were not performed.
When applying this process to all \enface slices across an entire volume, the total processing time was around 281.6 s/volume (i.e., 4.7 min/volume).

\begin{table}
	\centering
	\caption{Time required for each step from data acquisition to the generation of a speckle-reduced image in FF-SSOCT.}
	\label{tab:processing_time}
	\begin{tabular}{lccc}
		\hline
		Processing step & Time (s) & Percentage (\%) \\
		\hline
		Data acquisition & 0.25 & 0.02 \\
		Data saving & 16.0 & 0.98 \\
		Spectral rescaling + FFT & 94.5 & 5.79 \\
		High-frequency noise filtering & 104.7 & 6.41 \\
		Computational refocusing & 1151.9 & 70.55 \\
		Speckle reduction (proposed) &\textbf{ 281.6} &\textbf{ 17.25} \\
		\hline
		Total & 1632.7 & 100.00 \\
		\hline
	\end{tabular}
\end{table}
In addition to this core process, the complete imaging and post-processing workflow consists of several sequential stages from data acquisition to final image generation.
The time required for each step is summarized in Table \ref{tab:processing_time}.
As shown in the table, the total post-processing time per OCT volume accumulates to 1,632.7 s/volume (i.e., 27.2 min/volume).
The primary computational bottleneck is computational refocusing (excluding manual parameter selection), which consumes 70.6\% of the total timeline.
This heavy refocusing step is strictly necessary for FF-SSOCT imaging due to the system's narrow physical depth of focus, which is approximately 16 \um.

In summary, the proposed speckle reduction method accounts for only 17.25\% of the overarching processing profile and expands the total pipeline timeline by a modest 20.8\%.
Given that it requires only post-acquisition signal processing without any specialized hardware modifications or data acquisition protocols, this computational overhead is reasonable.

For the optimization of shift parameters (described in Section \ref{subsec:Resolution preservation analysis}), Fourier-domain up-sampling (zero-padding) was introduced to accommodate sub-pixel shifts.
However, experimental results revealed that optimal speckle-reduction performance is achieved only with integer shifts, specifically, a maximum lateral shift of 8 pixels with a 1-pixel step size.
This indicates that up-sampling and its execution time can be omitted in practical applications.

\subsection{Comparison with other speckle-reduction methods}
\subsubsection{Comparison with frame averaging}
\begin{figure}
	\centering
	\includegraphics[width=4.51in]{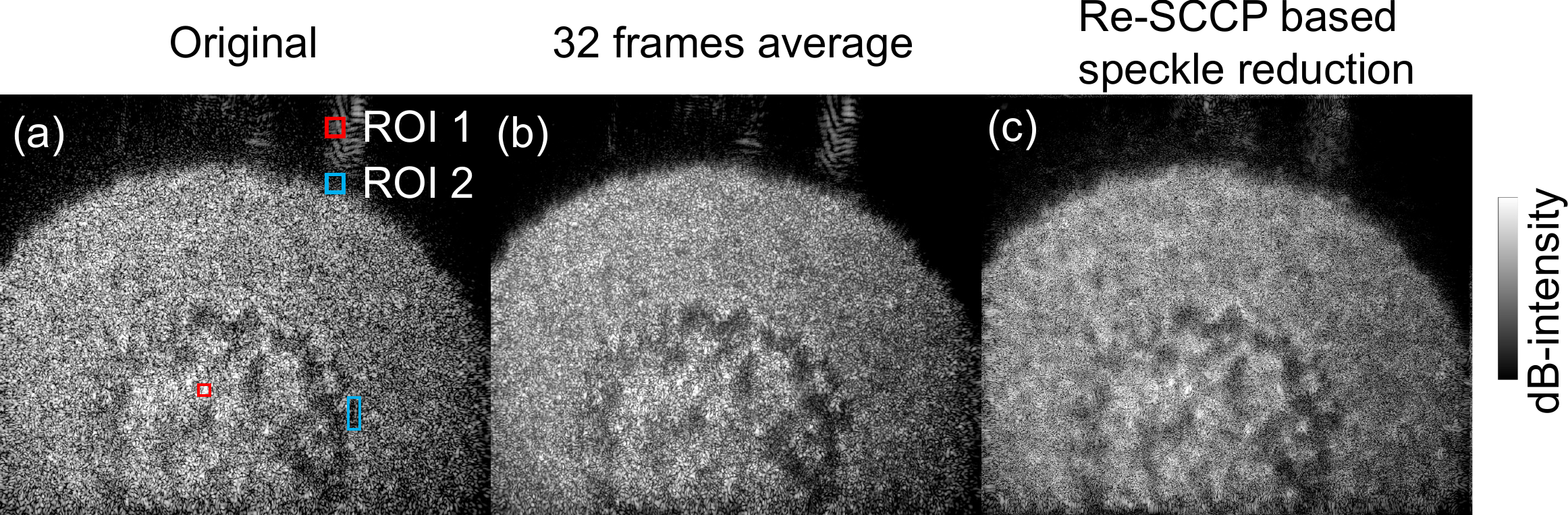}
	\caption{Comparison of speckle reduction performance for a sample cultured for 8 days without treatment.
		(a) Original FF-SSOCT \enface image with two selected regions of interest (ROI1 and ROI2).
		(b) Speckle-reduced image obtained by conventional 32-frame averaging.
		(c) Speckle-reduced image obtained using the proposed SCCP-based method, with a maximum shift amount (MSA) of 7.68 \um and averaging over 145 real SCCP images.
		Compared with frame averaging, the SCCP-based method more effectively suppresses speckle noise while revealing clearer microstructural details and improved image contrast.}
	\label{fig:ComparisonFA}
\end{figure}
Frame averaging is the most widely adopted speckle reduction strategy in clinical OCT imaging \cite{gotzinger2011speckle, alonso2011speckle, tan2012speckle}.
Here, we directly compared our optimal SCCP protocol against a 32-frame averaging approach using an \enface slice of the same tumor spheroid sample shown in Fig.\@ \ref{fig:OptimalSpheroid} (at a different depth).
The 32 volumes were acquired sequentially in 8.02 s, and are equivalent to the volume sequence used to compute dynamic OCT \cite{tateno2026dynamic, Komeda2026BOE}.

The comparison of results reveal distinct advantages for the proposed method, as follows.
As shown in Fig.\@ \ref{fig:ComparisonFA}, the SCCP-based method (MSA = 7.68 \um, NUI = 145) more effectively suppresses granular noise and reveals clear microstructural features that remain partially obscured in the 32-frame-averaged image.
In addition, the SCCP method achieves a significantly higher CNR (2.79 dB) than both the original image (-0.44 dB) and the frame-averaged image (-0.11 dB), where the CNR values were computed using the two ROIs indicated in the figure.

In addition, conventional frame averaging methods suffer from a fundamental mismatch when applied to \invitro (and also \exvivo) conditions and full-field swept-source OCT.
First, frame averaging relies on the temporal decorrelation of speckle patterns.
Since the temporal decorrelation is mainly caused by involuntary sample motions, this method is highly effective for \invivo imaging, such as clinical retinal imaging.
On the other hand, \invitro and \exvivo samples (e.g., tumor spheroids) are relatively stable, which causes a high inter-frame correlation of speckle patterns and makes the frame-averaging-based speckle reduction inefficient.

Second, frame averaging is traditionally optimized for B-scan-oriented imaging protocols, such as cross-sectional scans in point-scanning OCT.
A single wavelength scan of FF-SSOCT captures not a single B-scan but a volumetric dataset.
In such a protocol, a large number of volumes (not just B-scans) must necessarily be acquired for frame averaging, which imposes a heavy burden on memory and image processing.

\subsubsection{Comparison with hardware-based methods}
While several alternative speckle reduction techniques have been developed, they generally rely on specialized acquisition hardware.
Angular compounding methods use multiple OCT images obtained at multiple angles \cite{desjardins2006speckle, zhao2020angular}.
Although this greatly reduces the appearance of speckle, it requires specialized hardware that enables multi-angle image detection.

Polarization-compounding methods, such as that in Ref.\@ \cite{storen2004comparison}, can also reduce speckle, but they require polarization diversity detection and/or polarization multiplexing of the incident polarization.
In addition, since the degree of freedom of polarization is only two, at most two fully uncorrelated speckle patterns can be generated.
This theoretically limits the performance of the method.

Finally, wavelength compounding methods can also reduce speckle.
However, they may sacrifice axial resolution because each OCT image to be averaged is generated by a narrow-windowed spectral interference signal, or they require multiple light sources in different wavelength bands \cite{pircher2003speckle}.

In addition, the hardware extensions required for polarization- and wavelength-compounding methods are not well suited for FF-SSOCT, since they require multiple high-speed cameras and/or multiple wavelength-sweeping light sources.
In contrast, the real-SCCP-based method operates entirely with post-acquisition signal processing.
It requires only a single volumetric acquisition, demands no hardware modifications, and preserves both lateral and axial image resolution.
This makes the method highly suitable for FF-SSOCT.

\section{Conclusion}
An imaging-model-based numerical speckle reduction method for FF-SSOCT was presented.
PSF phantom experiments confirm that the proposed method preserves lateral resolution.
The application to the \invitro tumor spheroid and postmortem zebrafish eye demonstrated improved image observations, and quantitative image quality metrics demonstrated the good speckle-reduction performance of the real-SCCP-based method.
According to its software-only implementation, resolution-preserving nature, and remarkable speckle reduction performance, it can be concluded that our proposed speckle reduction method is a practical and effective way to improve the image quality of OCT.

\begin{backmatter}
	\bmsection{Funding}
Japan Science and Technology Agency (JPMJCR2105, JPMJSP2124);
Japan Society for the Promotion of Science (26H02451, 21H01836, 22K04962).

	\bmsection{Disclosures}
	Wang, Makita, Tateno, Komeda, Bao, Yasuno: Nidek(F), Sky Technology(F), Panasonic(F), Nikon(F), Santec(F), Kao Corp.(F), Topcon(F).(F);
	Furukawa, Matsusaka: None;
	Kobayashi: None;
	Tateno is currently employed by Kowa Company, Ltd.
	
	\bmsection{Data availability} Data underlying the results presented in this paper are not publicly available at this time but may be obtained from the authors upon reasonable request.
	
\end{backmatter}

\bibliography{SpeckleReductionPaper}

\end{document}